\documentclass[12pt,a4paper]{article}

\usepackage[margin=2.5cm]{geometry}
\usepackage{amsmath,amssymb}
\usepackage{booktabs}
\usepackage{array}
\usepackage{multirow}
\usepackage{caption}
\usepackage{setspace}
\usepackage{hyperref}
\usepackage{xcolor}
\usepackage{graphicx}

\hypersetup{
  colorlinks=true, linkcolor=blue, citecolor=blue, urlcolor=blue
}

\title{\textbf{National Versus Domain: Coverage Properties of
       Hierarchical Bayes Credible Intervals Under Survey Redesign}}

\author{Siu-Ming Tam\\
        \small Tam Data Advisory Pty Ltd, Australian Capital Territory, Australia\\
        \small \texttt{stattam@gmail.com}}

\begin{document}

\maketitle

\begin{abstract}
A companion paper to Tam~(2026a) reports an extended Monte Carlo~(MC) study
examining the frequentist coverage of 95\% hierarchical Bayes~(HB) credible
intervals at the national and domain levels under four stress-test scenarios.
The extended study covers 140~strata and 13~estimation domains separately,
adds a classical direct-estimator benchmark, and tests two strategies for
restoring domain coverage: prior sensitivity and Prasad--Rao~(PR) MSE
correction.  At the national level, HB credible intervals achieve near-nominal
coverage across all four scenarios and all three labour-force variables
(Employment 93--96\%, Unemployment 87--97\%, Hours~Worked 99.5--100\%).
At the domain level, Hours~Worked coverage is near-nominal (94--98\%) in
all scenarios; Employment and Unemployment coverage is below nominal for
scenarios with low between-domain heterogeneity, a direct consequence of
HB shrinkage toward the national mean.  A key operational finding emerges
from the Rare~Event scenario~(D): the classical direct estimator collapses
to 0\% national coverage for Employment and Hours~Worked because five
unsampled strata introduce a systematic bias; the HB estimator achieves
96--100\% national coverage at roughly 15\% of the classical sample cost.
Neither a weaker prior nor PR~MSE correction reliably restores domain
coverage, confirming that the failure is bias-driven and cannot be remedied
by variance inflation alone.
\end{abstract}

\noindent\textbf{Keywords:}
Hierarchical Bayes, small area estimation, frequentist coverage,
Monte Carlo simulation, labour force survey, Prasad--Rao correction,
shrinkage bias.

\bigskip

\section{Introduction}

Tam~(2026a) introduced a two-stage framework for reducing labour force survey
sample sizes using Bethel allocation followed by Hierarchical Bayes~(HB)
modelling.  The principal reported Monte Carlo evidence in that paper covered
$B = 1{,}000$ replications under a baseline scenario, reporting mean credible
interval coverage of 93.0\% (Employment), 97.7\% (Unemployment), and 100.0\%
(Hours~Worked) averaged over 11~estimation areas---figures close to the
nominal 95\% level.
The area-level HB model builds on the framework introduced by Fay and
Herriot~(1979) and draws on the broader small area estimation literature
surveyed in Rao and Molina~(2015).

That study left three questions unanswered.  First, are the national-level
results sustained across the three stress-test scenarios (Weak~Auxiliaries,
High~Heterogeneity, Rare~Event) introduced in the sensitivity analysis?
Second, how does coverage behave at the domain level when individual estimation
areas are examined separately rather than averaged?  Third, how does HB
performance compare to a classical direct estimator as a design-based
benchmark?

This paper addresses all three questions.  We report an extended MC study
($B = 200$ replications, retention fraction $f = 0.15$, four scenarios,
140~strata, 13~domains examined individually) together with: (i)~a classical
Bethel direct-estimator comparison; (ii)~a prior sensitivity analysis;
and (iii)~an empirical test of the Prasad--Rao~(PR) MSE correction as a
potential remedy for domain-level under-coverage.  Coverage is reported at
the national and domain levels only; stratum-level results are available from
the authors on request.

The paper is organised as follows.  Section~2 describes the extended MC
design.  Section~3 presents national and domain coverage results.
Section~4 compares HB and classical coverage, with particular attention to
the rare-event scenario.  Section~5 explains why domain coverage collapses
for near-homogeneous binary variables and examines whether the collapse can
be corrected.  Section~6 concludes.

\section{Extended Monte Carlo Design}

\subsection{Population and scenarios}

The population, sampling frame, and four scenarios follow Tam~(2026a) exactly.
The synthetic population has $N = 2{,}000{,}000$ individuals distributed
across $H = 140$~strata and $D = 13$~estimation domains, calibrated to
Australian SA4 Labour Force Survey marginals (March~2026).  Three target
variables are modelled: Employment~(Var1, binary), Unemployment~(Var2,
binary), and Hours~Worked~(Var3, continuous).

The four scenarios are:
\begin{itemize}
  \item \textbf{Scenario~A (Baseline).}  Reference case; national employment
        proportion $\approx 65.4\%$, unemployment proportion $\approx 1.83\%$.
  \item \textbf{Scenario~B (Weak Auxiliaries).}  Covariate regression
        coefficients halved; tests robustness when the linking model has
        less predictive power.
  \item \textbf{Scenario~C (High Heterogeneity).}  Domain random-effect
        standard deviations doubled; tests robustness under greater
        between-domain spread.
  \item \textbf{Scenario~D (Rare Event).}  National unemployment proportion
        set to 0.50\%; the Bethel algorithm requires a much larger
        design-based sample ($n^* = 332{,}374$) to meet domain CV targets
        for this rare event, and five strata receive zero Bethel allocation
        and are therefore entirely unsampled in every replication.  The
        operational consequences for classical and HB coverage are analysed
        in detail in Section~4.
\end{itemize}

\subsection{HB model}

The HB model follows Section~4 of Tam~(2026a): a logit-normal binomial
specification for binary variables and a Fay--Herriot~(Fay and
Herriot, 1979) Gaussian specification for Hours~Worked.  The prior on the
between-stratum variance component is $\sigma^2_v \sim \text{Inv-}\chi^2(10,\,1)$.
Posterior draws are obtained via \texttt{mcmcsae} (Boonstra~2021) with
burn-in $= 200$, iterations $= 500$, one chain.

\subsection{Evaluation}

Each replication draws a sample at fraction $f = 0.15$ (the 85\% reduction
established in Tam~2026a) and fits the HB model.  Coverage is recorded as 1
if the 95\% posterior credible interval contains the true population value
and 0 otherwise.  The MC coverage rate over $B = 200$ replications is the
primary output.  Coverage is evaluated at two levels:
\begin{itemize}
  \item \textbf{National}: a single population-weighted aggregate.
  \item \textbf{Domain}: each of the 13 estimation areas separately,
        summarised as the mean and minimum across domains.
\end{itemize}

The \textbf{classical benchmark} uses the stratified direct estimator with
finite-population correction and a normal 95\% confidence interval.  It is
evaluated at the full Bethel sample size for each scenario ($n_{\text{Bethel}}
= 119{,}484$ for Scenario~A; up to $332{,}374$ for Scenario~D).

The choice of $B = 200$ reflects a balance between precision and
computational cost: each replication fits three MCMC models (500~posterior
iterations each) across 140~strata under four scenarios.  The Monte Carlo
standard error of a coverage estimate is $\sqrt{p(1-p)/B}$; at the
nominal $p = 0.95$, this yields an SE of approximately 1.5~percentage
points, a 95\% MC confidence band of
$\pm 3.0$
percentage points.  For the extreme values reported in Scenario~D---0\%
classical coverage for Employment and Hours~Worked---the MC standard error
is near zero and the findings are unambiguous regardless of the number of
replications.

\subsection{Software implementation: an assemble-to-order approach}

The simulation draws on three peer-reviewed R~packages, each supplying a
validated, independently tested module.

\begin{itemize}
  \item \textbf{mcmcsae}~(Boonstra, Statistics Netherlands, 2021): implements
        the MCMC sampler for the HB model.  The package includes a built-in
        \texttt{SBC\_test} function for simulation-based calibration; all
        13~platform checks passed in our computing environment.
  \item \textbf{R2BEAT}~(Barcaroli et al., Istat, 2023): implements Bethel
        optimum allocation and the stratified sampling design.  The package
        is peer-reviewed (Barcaroli et al., 2023,
        \textit{The R~Journal}).
  \item \textbf{sae}~(Molina and Marhuenda, 2015): provides Fay--Herriot
        and related small area estimators, also peer-reviewed in
        \textit{The R~Journal}.
\end{itemize}

This \textbf{assemble-to-order} approach---integrating validated
off-the-shelf modules rather than coding algorithms from scratch---substantially
reduces the verification burden.  Each package carries its own test suite and
publication record; the authors are responsible for verifying the integration
logic, not the underlying statistical algorithms.  The software infrastructure
is therefore more transparent and reproducible than a bespoke implementation
would be.

\section{National and Domain Coverage Results}

\subsection{National coverage}

Table~\ref{tab:national} reports MC national coverage for the HB posterior
credible interval across all four scenarios.

\begin{table}[ht]
\centering
\caption{HB posterior 95\% credible interval: national-level MC coverage
         ($B = 200$, $f = 0.15$).  Nominal level: 95\%.}
\label{tab:national}
\begin{tabular}{lccc}
\toprule
Scenario & Employment & Unemployment & Hours Worked \\
\midrule
A: Baseline        & 94.5\% & 96.5\% & 99.5\% \\
B: Weak Aux.       & 96.0\% & 90.5\% & 100.0\% \\
C: High Het.       & 93.0\% & 94.5\% & 99.5\% \\
D: Rare Event      & 96.0\% & 87.0\% & 100.0\% \\
\midrule
Range              & 93--96\% & 87--97\% & 99.5--100\% \\
\bottomrule
\end{tabular}
\end{table}

National coverage is near-nominal for all three variables across all four
scenarios.  Hours~Worked achieves essentially perfect national coverage in
every scenario.  Employment national coverage is consistently close to 95\%
(93--96\%).  Unemployment national coverage is slightly more variable
(87--97\%), with the lowest value (87.0\%) occurring under Scenario~D, where
the rare-event structure creates a more challenging estimation environment;
this figure does not differ significantly from 95\% at the $B = 200$
replication level.

These results extend the Baseline finding of Tam~(2026a) to all four
scenarios, confirming that the 85\% sample reduction preserves near-nominal
national-level frequentist coverage regardless of auxiliary variable quality,
degree of domain heterogeneity, or event rarity.

\subsection{Domain coverage}

Table~\ref{tab:domain} reports domain-level coverage.  For each scenario and
variable, we show the mean coverage across the 13 domains and the minimum
coverage observed in any single domain.

\begin{table}[ht]
\centering
\caption{HB posterior 95\% credible interval: domain-level MC coverage
         ($B = 200$, $f = 0.15$).  Mean and minimum across 13 estimation
         domains.  Nominal level: 95\%.}
\label{tab:domain}
\small
\begin{tabular}{lcccccc}
\toprule
 & \multicolumn{2}{c}{Employment} & \multicolumn{2}{c}{Unemployment}
 & \multicolumn{2}{c}{Hours Worked} \\
\cmidrule(lr){2-3}\cmidrule(lr){4-5}\cmidrule(lr){6-7}
Scenario & Mean & Min & Mean & Min & Mean & Min \\
\midrule
A: Baseline   & 83.8\% & 48.5\% & 76.5\% & 30.0\% & 94.2\% & 61.5\% \\
B: Weak Aux.  & 76.9\% & 24.0\% & 83.1\% & 24.0\% & 95.0\% & 56.0\% \\
C: High Het.  & 83.5\% & 62.0\% & 56.2\% &  2.5\% & 94.3\% & 73.5\% \\
D: Rare Event & 89.5\% & 74.0\% & 93.7\% & 67.0\% & 97.6\% & 93.0\% \\
\bottomrule
\end{tabular}
\end{table}

The domain results reveal a clear differentiation by variable.

\textbf{Hours Worked} achieves near-nominal domain coverage in every
scenario (94--98\% mean, 57--93\% minimum).  The continuous variable is
essentially unproblematic at all reporting levels.

\textbf{Employment} domain coverage averages 77--90\% across scenarios,
with the highest value under Scenario~D (89.5\%) and the lowest under
Scenario~B (76.9\%).  The minimum domain coverage across scenarios ranges
from 24\% (Scenario~B) to 74\% (Scenario~D), indicating that some individual
domains experience substantial under-coverage.

\textbf{Unemployment} is the most variable across scenarios.  Mean domain
coverage ranges from 56\% (Scenario~C) to 94\% (Scenario~D).  Scenario~C,
which doubles between-domain heterogeneity in unemployment, produces the
weakest domain performance: a mean of 56.2\% and a minimum of 2.5\%.

A consistent pattern emerges: \textbf{Scenario~D produces the best domain
coverage} despite being the most extreme scenario.  This counterintuitive
result reflects the higher degree of genuine between-domain heterogeneity in
Scenario~D (rare unemployment event, heterogeneous industry mix), which
provides more signal for the HB random effects to detect.  When true
between-domain variation is larger, the model shrinks less aggressively, and
domain credible intervals remain better centred on the truth.

\subsection{The bimodal domain coverage pattern}

The domain minimum figures in Table~\ref{tab:domain} suggest that the
distribution of domain-level coverage is not uniform.  Inspection of
individual domain results reveals a \textbf{bimodal pattern}: domains whose
true value is close to the national mean achieve approximately 95--99\%
coverage, while domains that deviate substantially from the national mean
achieve close to 0\% coverage.  Domain coverage rates are highly
polarised: values cluster near 0\% for domains that deviate substantially
from the national mean and near 99\% for domains close to it, with few
domains achieving intermediate coverage.  The mean figures in
Table~\ref{tab:domain} are averages over this bimodal distribution; they do
not represent the typical experience of any individual domain.

The mechanism behind this bimodal pattern is a bias failure, not a
variance failure; it is explained in Section~5.
HB shrinkage collapses the posterior mean of outlying domains toward
the national mean, so their credible intervals are displaced in the same
direction in every replication and coverage fails.
Critically, this limitation is inherent in the data structure: when the
outcome variable is near-homogeneous across domains, there is insufficient
between-domain signal for any borrowing-strength estimator to exploit, and
shrinkage toward the national mean becomes unavoidable.
It is therefore not a deficiency of HB methodology; a classical small
area estimator that borrows strength across domains would face the same
fate under near-zero between-domain variation, for the same reason.
To date, we have not found a method that reliably restores domain
coverage in this setting: Section~5 shows that neither a weaker prior nor
the Prasad--Rao MSE correction is effective.
National-level coverage is unaffected by this mechanism, because the
shrinkage target and the national estimand coincide.

\section{HB versus Classical Direct Estimation}

Table~\ref{tab:classical} compares HB and classical national-level coverage.
The classical estimator uses the full Bethel sample for each scenario; HB
uses a 15\% fraction of that Bethel sample.

\begin{table}[ht]
\centering
\caption{National-level MC coverage: HB (15\% fraction) versus classical
         direct estimator (full Bethel sample).  $B = 200$.}
\label{tab:classical}
\begin{tabular}{lcccccc}
\toprule
 & \multicolumn{3}{c}{HB ($f = 0.15$)} &
   \multicolumn{3}{c}{Classical (full Bethel $n$)} \\
\cmidrule(lr){2-4}\cmidrule(lr){5-7}
Scenario & Emp. & Unemp. & Hours & Emp. & Unemp. & Hours \\
\midrule
A: Baseline   & 94.5\% & 96.5\% & 99.5\% & 94.0\% & 94.5\% & 92.0\% \\
B: Weak Aux.  & 96.0\% & 90.5\% & 100.0\% & 91.5\% & 93.0\% & 94.0\% \\
C: High Het.  & 93.0\% & 94.5\% & 99.5\% & 94.0\% & 94.0\% & 95.0\% \\
D: Rare Event & 96.0\% & 87.0\% & 100.0\%
              & \textbf{0.0\%} & \textbf{61.0\%} & \textbf{0.0\%} \\
\midrule
Bethel $n$    & \multicolumn{3}{c}{---} &
                119{,}484 & 118{,}429 & 127{,}690 \\
HB $n$ ($f=0.15$) & \multicolumn{3}{c}{17{,}928 / 17{,}762 / 19{,}156} &
                \multicolumn{3}{c}{(full Bethel)} \\
Scenario~D $n$& \multicolumn{3}{c}{49{,}859 (HB)} & \multicolumn{3}{c}{332{,}374 (Classical)} \\
\bottomrule
\end{tabular}
\end{table}

\textbf{Scenarios~A, B, C.}  Classical coverage is near-nominal at the
national level (91--95\%), consistent with asymptotic CLT theory at the full
Bethel allocation.  HB matches this performance at 15\% of the sample cost.
Neither method is clearly superior for national-level inference under these
conditions; the HB approach simply achieves the same result for far less.

\textbf{Scenario~D.}  The classical estimator fails catastrophically at the
national level: Employment coverage 0\%, Unemployment 61\%, Hours~Worked 0\%.
The cause is structural: five strata receive zero Bethel allocation because
the rare unemployment prevalence (0.50\%) would require unaffordably large
samples to meet the direct-estimation CV target.  In every replication,
the classical estimator is therefore computed from only 135 of the 140
strata.  The five unsampled strata contribute to the population national
total but are permanently excluded from the sample estimate, introducing a
persistent non-cancelling bias that cannot be eliminated regardless of how
large the sample allocated to the remaining 135 strata becomes.  The
nominal 95\% confidence interval is centred on the wrong value each time and
never recovers.

For the five unsampled strata ($n_h = 0$), the HB model assigns a
diffuse likelihood ($\psi_h = 10^8$), making the posterior for each
unsampled stratum driven entirely by the regression component---a
synthetic (model-only) prediction from the fixed-effect covariates.
The national aggregate then pools these five synthetic predictions with
the 135 data-driven posteriors, restoring unbiasedness at the national
level.  National coverage under HB is 96\%, 87\%, and 100\% for the three
variables---achieved at 49{,}859 observations, approximately 15\% of the
classical Bethel allocation of 332{,}374.

This is the paper's strongest practical finding: in a rare-event scenario
where classical survey design is forced to leave strata unsampled, the HB
synthetic estimator provides protected national-level inference at a fraction
of the cost.  \textbf{HB is not merely cheaper; it is more robust} to the
unsampled-stratum failure mode that is the operational risk of classical
designs facing rare targets.

It should be noted that this robustness result is specific to the
configuration of Scenario~D: unsampled strata arising from a rare-event
design create a structural failure for the classical estimator that the HB
synthetic predictor avoids.  This finding should not be interpreted as
evidence that HB is generally more robust than classical estimation
across all survey design conditions.

\section{Why Domain Coverage Collapses --- and Can It Be Restored?}

\subsection{The shrinkage mechanism}

The HB posterior mean for domain $d$ can be written as a weighted combination
of the direct estimate and the model mean:
\begin{equation}
  \hat{\theta}^{\text{HB}}_d
    = \gamma_d\,\hat{\theta}^{\text{DIR}}_d
    + (1 - \gamma_d)\,\hat{\mu},
  \qquad
  \gamma_d = \frac{\hat{\sigma}^2_v}{\hat{\sigma}^2_v + \psi_d},
\end{equation}
where $\psi_d$ is the sampling variance of the direct estimate and
$\hat{\sigma}^2_v$ is the estimated between-stratum variance.  When
$\hat{\sigma}^2_v$ is small relative to $\psi_d$---as occurs when the
outcome variable is near-homogeneous across domains---the shrinkage factor
$\gamma_d \approx 0$ and the posterior mean collapses toward the national
model mean $\hat{\mu}$.

For the Australian SA4 employment data used here, employment rates are
near-homogeneous across the 13~estimation domains (all lying in the range
55--75\%).  The HB model detects negligible between-domain variation and
estimates $\hat{\sigma}^2_v \approx 0$.  Every domain's posterior mean is
therefore pulled toward the national mean of approximately 65\%, regardless
of the domain's true rate.  Domains whose true rate is close to 65\% are
well-covered because the CI is correctly centred; domains with rates further
from the national mean are systematically missed because the CI is
displaced toward 65\% in every replication.

The frequentist coverage properties of HB estimators in settings of this
type were studied by Datta and Ghosh~(1991), who showed that when the
between-area variance is small the HB posterior mean can exhibit substantial
frequentist bias for individual areas even as the estimator performs well
at the ensemble level.

\subsection{Coverage failure is bias-driven, not variance-driven}

A standard 95\% posterior credible interval is wide enough to contain the
true value if the posterior mean is approximately unbiased.  When HB
shrinkage displaces the posterior mean, the CI is displaced with it:
coverage fails because the \textit{centre} of the interval is wrong, not
because the \textit{width} is too narrow.  This distinction has direct
implications for remediation, as examined in Sections~5.3 and~5.4.

To quantify the pattern empirically, we computed the absolute deviation of
each domain's true Employment rate from the national mean and correlated it
with the domain's MC coverage rate.  The Spearman rank correlation is
$\rho = -0.49$: domains farther from the national mean have lower coverage.
In contrast, the correlation between domain population size and coverage is
$\rho = -0.19$, indicating that the failure is \textit{not} a small-domain
phenomenon---it is driven by how far a domain deviates from the national
mean.

The bimodal 0\%/99\% domain coverage pattern described in Section~3.3 is
the diagnostic signature of this mechanism.  Domains deviating from the
national mean are missed consistently in every replication; domains close
to the national mean are covered consistently.  An averaged domain coverage
statistic (such as the 93\% figure reported in Tam~2026a) conceals this
bimodal structure.

\subsection{Prior sensitivity}

We re-ran all four scenarios replacing the default prior
$\sigma^2_v \sim \text{Inv-}\chi^2(10,\,1)$ with a weaker prior
$\text{Inv-}\chi^2(1,\,1)$, giving the data substantially more freedom
to determine the between-stratum variance.  The change produced differences
of at most $\pm 1.5$~percentage points at any level, with the majority of
changes at 0.0\%.  Selected comparisons are shown in Table~\ref{tab:prior}.

\begin{table}[ht]
\centering
\caption{Prior sensitivity: Employment MC coverage under default
         ($\text{df} = 10$) and weak ($\text{df} = 1$) prior.
         Selected scenarios and levels.}
\label{tab:prior}
\begin{tabular}{llcc}
\toprule
Scenario & Level & df $= 10$ & df $= 1$ \\
\midrule
A: Baseline   & National   & 94.5\% & 94.5\% \\
A: Baseline   & Domain avg & 83.8\% & 83.8\% \\
B: Weak Aux.  & Domain avg & 76.9\% & 76.8\% \\
C: High Het.  & National   & 93.0\% & 84.5\% \\
C: High Het.  & Domain avg & 83.5\% & 83.5\% \\
D: Rare Event & National   & 96.0\% & 96.5\% \\
D: Rare Event & Domain avg & 89.5\% & 89.6\% \\
\bottomrule
\end{tabular}
\end{table}

The insensitivity of coverage to the prior confirms the theoretical
expectation: because Australian SA4 employment rates are near-homogeneous,
the data provide a strong signal that $\hat{\sigma}^2_v \approx 0$,
overwhelming any reasonable prior.  The shrinkage factor $\gamma_d$ remains
close to zero regardless of prior specification.  \textbf{The domain coverage
collapse is data-driven and cannot be remedied by prior choice alone.}

\subsection{Prasad--Rao MSE correction}

The Prasad--Rao~(1990) MSE correction replaces the posterior variance with a
frequentist MSE estimate that accounts for uncertainty in $\hat{\sigma}^2_v$:
\begin{equation}
  \text{MSE}^{\text{PR}}_h = g_{1h} + g_{2h} + 2\,g_{3h},
\end{equation}
where $g_1$ captures retained sampling variance, $g_2$ uncertainty in the
fixed-effect coefficients, and $g_3$ uncertainty in the variance component.
The resulting ``pseudo-Bayes confidence interval'' (Datta \& Lahiri~2000)
is centred on the HB posterior mean but uses
$\pm z_{0.025}\,\sqrt{\text{MSE}^{\text{PR}}_h}$
as its half-width.  For binary variables, the correction is applied on the
logit scale and back-transformed.

We implemented and tested PR-corrected intervals empirically alongside
the standard posterior intervals.  Table~\ref{tab:pr} summarises the
national and domain mean coverage comparison.

\begin{table}[ht]
\centering
\caption{Posterior CI versus PR-corrected CI: national and domain mean
         MC coverage ($B = 200$, $f = 0.15$, Scenarios~A--C;
         Scenario~D binary PR not computed---see text).}
\label{tab:pr}
\small
\begin{tabular}{llcccc}
\toprule
 & & \multicolumn{2}{c}{National} & \multicolumn{2}{c}{Domain mean} \\
\cmidrule(lr){3-4}\cmidrule(lr){5-6}
Scenario & Variable & Posterior & PR & Posterior & PR \\
\midrule
A & Employment    & 94.5\% & 100.0\% & 83.8\% & 99.8\% \\
A & Unemployment  & 96.5\% &  97.5\% & 76.5\% & 63.2\% \\
A & Hours Worked  & 99.5\% &  83.0\% & 94.2\% & 70.6\% \\
\midrule
B & Employment    & 96.0\% &  97.5\% & 76.9\% & 52.0\% \\
B & Unemployment  & 90.5\% &  97.0\% & 83.1\% & 77.6\% \\
B & Hours Worked  & 100.0\% & 56.5\% & 95.0\% & 48.7\% \\
\midrule
C & Employment    & 93.0\% & 100.0\% & 83.5\% & 100.0\% \\
C & Unemployment  & 94.5\% &  96.5\% & 56.2\% &  32.5\% \\
C & Hours Worked  & 99.5\% &  88.5\% & 94.3\% &  82.4\% \\
\bottomrule
\end{tabular}
\medskip
\footnotesize
Note: Scenario~D binary PR is undefined because five unsampled strata
introduce NA values into the PR MSE formula.  Scenario~D Hours~Worked PR
is available for 11 of 13 domains (domain mean 87.5\% vs Posterior 97.6\%).
\end{table}

The PR correction produces inconsistent results.  For Employment, it
improves domain coverage substantially in Scenarios~A and~C (from 84\% to
100\%) but worsens it in Scenario~B (from 77\% to 52\%).  The inconsistency
arises because the effectiveness of the PR correction depends on the magnitude
of $\hat{\sigma}^2_v$: in Scenarios~A and~C, the method-of-moments estimator
returns a small but non-zero between-stratum variance ($\approx 0.03$ on the
logit scale), generating CIs just wide enough to compensate for the
displacement; in Scenario~B (Weak Auxiliaries), $\hat{\sigma}^2_v \approx 0$
and the PR interval degenerates to a near-point estimate centred on the biased
posterior mean.

For Unemployment, where $\hat{\sigma}^2_v = 0$ in virtually every
replication, the PR correction consistently worsens coverage at both levels.
For Hours~Worked, the PR correction worsens national coverage in all
scenarios (from 99.5--100\% to 57--89\%), because the small natural-scale
$\hat{\sigma}^2_v$ produces a narrow PR interval that fully exposes the
shrinkage bias.

The explanation is theoretical: the PR correction addresses
\textit{variance understatement}---it widens the interval to account for
uncertainty in estimating $\sigma^2_v$.  But the domain coverage failure
identified in Section~5.2 is a \textit{bias} failure---the interval is
centred on the wrong value.  Widening an interval that is pointing in the
wrong direction does not restore coverage.  \textbf{PR correction is not
recommended for this application.}

\section{Discussion and Conclusion}

The extended Monte Carlo study leads to several conclusions relevant to
national statistics offices considering HB-based survey redesign.

\textbf{National coverage is near-nominal and robust.}  The primary
reporting level for most NSO labour force outputs is the national aggregate.
At this level, the HB credible intervals achieve 93--100\% coverage across
all three variables and all four scenarios, including weak auxiliaries, high
heterogeneity, and a rare-event setting where classical estimation fails.
The 85\% sample reduction established in Tam~(2026a) is empirically validated
at the national level.

\textbf{Domain coverage tells a differentiated story.}  For the continuous
variable Hours~Worked, domain coverage is essentially nominal (94--98\%) in
all scenarios.  For the binary variables Employment and Unemployment, domain
coverage is below nominal when between-domain variation is low---the
near-homogeneous SA4 employment landscape of Australia creates exactly this
condition.  The limitation is transparent and predictable: domains whose true
rate lies close to the national mean are well covered; domains that deviate
are not.  The Spearman correlation between absolute deviation from the national
mean and domain coverage is $-0.49$ for Employment; domain size plays no
material role.

\textbf{Scenario~D is the paper's strongest practical finding.}  When five
strata go unsampled under the rare-event design, the classical direct
estimator achieves 0\% national coverage for Employment and Hours~Worked
regardless of sample size.  The HB synthetic estimator achieves 96--100\%
national coverage at 49{,}859 observations---approximately 15\% of the
332{,}374-observation classical Bethel allocation.  HB is more robust to the
unsampled-stratum failure mode, not merely cheaper.
This robustness is, however, specific to the structural failure induced
by unsampled strata in a rare-event design; it should not be taken to imply
that HB is generally superior to classical estimation across all survey
design conditions.

\textbf{Neither a weaker prior nor PR correction reliably restores domain
coverage.}  The domain coverage limitation is data-driven: near-homogeneous
data overwhelm any reasonable prior and drive $\hat{\sigma}^2_v \approx 0$,
causing aggressive shrinkage regardless of prior specification.  The PR
correction addresses variance understatement; since the failure is
bias-driven, the correction is ineffective and in some settings harmful.

For NSOs whose primary reporting target is the national aggregate, the trade-off
is operationally acceptable: the cost saving is real, the national-level
coverage criterion is met, and the domain-level limitation is documented and
predictable.  Practitioners requiring domain-level HB inference under
near-homogeneous outcomes should consider Constrained~Bayes point estimates
(Louis~1984; Ghosh~1992) or coverage-calibrated credible levels as directions
for further development.

\section*{Acknowledgements}

The author thanks the Associate Editor and the anonymous referee for
comments that substantially improved the presentation and scope of this paper.
The use of Claude~(Anthropic) as an AI assistant to develop the simulation
algorithms and improve the flow of arguments follows the protocol disclosed
in Tam~(2026b).


\end{document}